# Rotating artificial gauge magnetic and electric fields

**V. E. Lembessis**[1], $A. Alqarni^1$, $S. Alshamari^1$, $A. Siddig^1 and O. M. Aldossary^{1,2}$

1. Department Physics Astronomy, King Saud University, P.O. Box 2455, Al Riyadh, Kingdom of Saudi Arabia 2. The National Center for Applied Physics, KACST, PO Box 6086, Riyadh 11442, Saudi Arabia

E-mail: `vlempesis@ksu.edu.sa`

**Abstract.** We consider the creation of artificial gauge magnetic and electric fields created when a two-level atom interacts with an optical Ferris wheel light field.These fields have the spatial structure of the optical Ferris wheel field intensity profile. If this optical field pattern is made to rotate in space then we have the creation of artificial electromagnetic fields which propagate in closed paths. The properties of such fields are presented and discussed

## 1. Introduction

The last years we have witnessed a fresh air blowing over condensed matter physics. This is due to major achievements in atomic physics that resulted from the advances in cooling and trapping of atomic motion [1]. Probably the most remarkable of these is the creation of optical lattices. These are synthetic lattices with a variety of symmetries in which we can trap atoms and create new synthetic forms of matter [2]. Such achievements have driven us towards the merging of cold atom physics with condensed matter physics and have paved the way for fascinating applications like the probing of exotic topological phases of matter [3].

Long time ago Feynman was the first who suggested that instead of trying to model quantum effects by conventional computers we can use simple and controllable quantum systems as quantum simulators for them [4]. There is a large number of condensed matter effects that are very hard to simulate on a classical computer. For example, high-temperature superconductivity and quantum magnetism. The difficulties are even stronger in cases where electrons are strongly interacting. Here Feynman's prophecy comes to rescue. Cold atoms are ideal quantum simulators for such cases since their configurations are highly tunable, i.e., all the interactions and parameters involved can be engineered to suit a given model [5], [6]. The Hubbard model and the superfluid Mott-insulator transition are two famous physical effects that that can be simulated with cold atoms in optical lattices [7], [8]. Another example, which is related to our work, is the generation of artificial Abelian and non-Abelian magnetic and electric fields when cold atoms interact with coherent light fields [9].

Artificial gauge fields are very important in areas like high-energy and condensed matter physics. The reason for this is that, since the atoms are electrically neutral, we cannot extend quantum simulations with neutral atoms to cases involving charged particles. It is therefore a prerequisite reason to construct such gauge fields for atoms.Today there are different schemes which may generate artificial magnetic and electric fields [2]. These schemes involve atoms interacting with properly structured light fields. The atoms may be in the free space or trapped in atomic traps or optical lattices.

In the case of free atoms, as it is the case in our work, the interaction of them with the light results in an atomic motion that mimics the dynamics of a charged particle inside a magnetic field with the emergence of a Lorentz-like force [9]. This is due to the Berry phase acquired by the particle when it moves along a closed path [10]. This phase has a geometrical nature, i.e., it does not depend on the duration needed to complete the trajectory. So if we wish to exhibit an artificial magnetism we must find situations where a neutral particle, for some reason, acquires a geometrical phase when it moves along a closed path C. To achieve this researchers have exploited the Berry phase effect in atom-light interactions [11], [12], [13]. In this case the atom-light coupling is represented by the so-called dressed states [14], which can vary on a short spatial scale (typically the wavelength of light) and the artificial gauge fields can be quite intense. If the atom at time $t = 0$ is prepared in a dressed state $|\chi(\mathbf{r}_0)\rangle$ and moves slowly enough, then it follows adiabatically the local dressed state $|\chi(\mathbf{r}_t)\rangle$. When the atom completes the trajectory C it returns to the dressed state $|\chi(\mathbf{r}_0)\rangle$ having acquired a phase factor that contains a geometric component. The quantum motion of the atom is formally equivalent to that of a charged particle in a static magnetic field. Such models have been studied for different beam configurations for two-level as well as three-level atoms [9]. It is very important to note that the emergence of these artificial fields requires a coherent interaction between the light fields and the atoms. Thus the interaction time must be limited in values $t < \Gamma^{-1}$, with $\Gamma$ being the spontaneous emission rate of the excited state.

In our work we consider the generation of artificial magnetic and electric gauge fields in the special case where a two-level atom interacts with an optical Ferris wheel. The optical Ferris wheel configuration is the result of the superposition of two co-propagating Gaussian-Laguerre (GL) laser beams with opposite helicities $\pm l$ [15]. This field has a transverse intensity pattern with $2l$ pedal-like bright regions forming, thus, a cylindrical lattice. This type of lattice was proposed for studying persistent currents and the Mott-insulator transition in a ring geometry [15]. In 2009, an experiment demonstrated the trapping of Rb atoms in micron traps of an optical ring lattice that formed by the interference of two co-propagating LG beams with opposite orbital angular number $l = \pm 1$ [17]. The optical Ferris wheel was proposed to trap two identical atoms which can operate as a nano-antenna [18] where parameters, such as $l$ and the beam waist, can be manipulated to control and drive directional emission of light. The idea of optical Ferris wheel can be generalised to a scheme of two counter-propagating beams with opposite helicities which gives the so called helical optical tubes of light. This configuration, either in the static or in the rotating version was proposed to act as a tool for the investigation of the differences between a classical and a quantum fluid [19], as a detector for the slow rotation of atom interferometers [20], as a cavity for atoms [21] or as an Archimedes' screw for atoms [22].

The reason why we focus on the optical Ferris wheel case is neither the relevant size of the artificial gauge fields, which are weak, nor their spatial structure which originates from the characteristic spatial structure of the intensity of the optical Ferris light field. It is the possibility of creating time-dependent artificial magnetic and electric fields which rotate in the azimuthal direction so they propagate in closed paths. To the best of our knowledge it is the first time that such an effect is reported.

It is well established that a light field with time-dependent intensity pattern or/and phase structure leads to the generation of time-dependent artificial gauge fields, or in other words artificial electromagnetic fields. We must recall that the generation of artificial gauge fields with free atoms requires that optical field, which interacts with the atom, has simultaneously spatial gradients in amplitude and phase. This is the reason why we cannot generate them with standing waves created from the interference of Gaussian beams. The phase of a GL beam, though, if we consider that it propagates along the positive $z-$ direction, contains a phase factor $\exp(ikz)$ and a phase factor $\exp(il\phi)$. We can thus create either an "axial" standing wave with two counter-propagating beams which carries orbital angular momentum since the resulting standing wave has a phase $\exp(il\phi)$, or an "azimuthal" standing wave with two beams of opposite helicity which propagates in the axial direction and has a phase factor $\exp(ikz)$. The later is the optical Ferris wheel light beam. The intensity pattern of a light field that is made by interfering two coherent light beams can move in space by introducing a frequency shift between the two laser beams. In the first case the pattern moves in the axial direction leading to the generation of artificial gauge electromagnetic fields that propagate in a straight line. In the case of the optical Ferris wheel light field the intensity pattern can be rotated leading to artificial electromagnetic gauge fields which propagate in closed paths. It is this rather unusual property which has led us to the study of the artificial gauge fields generated from the interaction of atom with the optical Ferris wheel light fields.

The structure of the paper is as follows: In Section 2 we review the basic properties of the electric field of the optical Ferris wheel light field and we derive the artificial gauge magnetic fields which are generated when this light field interacts with a typical two-level atom. In Section 3 we derive the artificial gauge magnetic and electric fields which are generated when the intensity pattern of the optical Ferris wheel light field rotates. In Section 4 we summarise our findings.

## 2. Artificial Gauge Fields generated by a static Ferris Wheel Light Field

As has been shown in [15], the optical Ferris field is created from the interference of two co-propagating Gaussian-Laguerre of wavelength $\lambda$ beams with opposite helicity $\pm l$. If we consider that the beams propagate along the $z-$ direction and are linearly polarized along the x-direction, then their electric field is given by $\mathbf{E}_{\pm l,p}(\mathbf{R}) = E_{lp}(r,z) \exp i\Theta_{\pm l,p}(r,z,\phi)\hat{\mathbf{x}}$ where:

$$E_{lp}(r,z) = E_0 \sqrt{\frac{p!}{(|l|! + p!)}} \left(\frac{r\sqrt{2}}{w(z)}\right)^{|l|} L_p^{|l|}\left(\frac{2r^2}{w^2(z)}\right) \exp\left(-\frac{r^2}{w^2(z)}\right), \qquad (1)$$

and

$$\Theta_{\pm l,p}(r.z,\phi) = kz \pm l\phi - (2p + |l| + 1)\tan^{-1}(z/z_R) + \frac{kr^2 z}{2(z^2 + z_R^2)} \qquad (2)$$

In the above relations we have that $w(z) = w_0\sqrt{1 + z^2/z_R^2}$, where $w_0$ is the beam waist and $z_R = \pi w_0^2/\lambda$ the Rayleigh range of the beam. The quantity $E_0$ is the electric field amplitude which corresponds to a Gaussian beam of the same power and beam waist while $L_p^{|l|}$ is the associated Laguerre polynomial. The two beams have equal frequency $\omega_L$. The total electric field of the resulting optical Ferris wheel has the following amplitude and phase:

$$E_F(r,z,\phi) = 2E_0 \sqrt{\frac{p!}{(|l|! + p!)}} \left(\frac{r\sqrt{2}}{w(z)}\right)^{|l|} L_p^{|l|}\left(\frac{2r^2}{w^2(z)}\right) \exp\left(-\frac{r^2}{w^2(z)}\right) \cos(l\phi), \qquad (3)$$

and

$$\phi_F(\mathbf{R}) = kz - (2p + |l| + 1)\tan^{-1}(z/z_R) + \frac{kr^2 z}{2(z^2 + z_R^2)}. \qquad (4)$$

Let's consider now a two-level atom with a transition frequency $\omega_0$ and an excited state spontaneous emission rate $\Gamma$. When the Ferris wheel light field interacts with this atom the interaction is determine by the frequency detuning $\delta = \omega_0 - \omega_L$ and the Rabi frequency:

$$\tilde{\Omega}(\mathbf{R}) = \Omega(\mathbf{R})\cos(l\phi), \qquad (5)$$

whith

$$\Omega(\mathbf{R}) = 2\Omega_0 \sqrt{\frac{p!}{(|l|! + p!)}} \left(\frac{r\sqrt{2}}{w(z)}\right)^{|l|} L_p^{|l|}\left(\frac{2r^2}{w^2(z)}\right) \exp\left(-\frac{r^2}{w^2(z)}\right), \qquad (6)$$

where the quantity $\Omega_0$ is the Rabi frequency corresponding to an interaction of the atom with a Gaussian beam of the same power and beam waist.

It is well established [14], that the interaction of the two-level atom with a coherent light field gives rise to two dressed states, namely

$$|\chi(\mathbf{r}_1(t))\rangle = \begin{pmatrix} \cos(\Theta(\mathbf{R}/2)) \\ \exp(\mathrm{i}\phi_F(\mathbf{R})) \sin(\Theta(\mathbf{R}/2)) \end{pmatrix}, \tag{7}$$

$$|\chi(\mathbf{r}_2(t))\rangle = \begin{pmatrix} -\exp(\mathrm{i}\phi_F(\mathbf{R})) \sin(\Theta(\mathbf{R}/2)) \\ \cos(\Theta(\mathbf{R}/2)) \end{pmatrix}, \tag{8}$$

with $\phi$ the position-dependent phase of the field and

$$\cos(\Theta(\mathbf{R})) = \frac{\delta}{\sqrt{\delta^2 + \tilde{\Omega}^2(\mathbf{R})}}. \tag{9}$$

Here $\delta$ is the detuning of the atomic transition from the frequency of the light and $\Omega(\mathbf{R})$ is the Rabi frequency given in Eq. (4). As has been shown [14], under adiabaticity conditions, we can get an artificial magnetic field given by

$$q\mathbf{B}(\mathbf{R}) = -\hbar\delta \frac{\tilde{\Omega}(\mathbf{R})}{(\delta^2 + \tilde{\Omega}^2(\mathbf{R}))^{3/2}} \overrightarrow{\nabla}(\tilde{\Omega}(\mathbf{R})) \times \overrightarrow{\nabla}(\phi_F(\mathbf{R})), \tag{10}$$

$$V(\mathbf{R}) = \frac{\hbar^2}{2M} \left[ \frac{\delta^2}{(\delta^2 + \tilde{\Omega}^2(\mathbf{R}))^2} (\overrightarrow{\nabla}\tilde{\Omega}(\mathbf{R}))^2 + \frac{\tilde{\Omega}^2(\mathbf{R})}{\delta^2 + \tilde{\Omega}^2(\mathbf{R})} \left( \overrightarrow{\nabla}\phi_F(\mathbf{R}) \right)^2 \right]. \tag{11}$$

In this case we may show that at the beam focus, $z = 0$, the magnetic field is given by $\mathbf{B}(\mathbf{R}) = B_r(\mathbf{R})\hat{r} + B_\phi(\mathbf{R})\hat{\phi}$, i.e. it has two components, one along the radial direction $r$ and another one along $\phi$ direction. The spatial dependence of these components is shown in Fig.2 for a specific choice of parameters if, also, we consider that our atoms have a fictitious charge equal to the charge of electron $q = |e|$. These components are given by:

$$B_r(\mathbf{R}) = \frac{\hbar k l \delta \tilde{\Omega}^2(\mathbf{R})}{q(\delta^2 + \tilde{\Omega}^2(\mathbf{R}))^{3/2}} \frac{\tan(l\phi)}{r}, \tag{12}$$

$$B_\phi(\mathbf{R}) = \frac{\hbar k l \delta \tilde{\Omega}^2(\mathbf{R})}{q(\delta^2 + \tilde{\Omega}^2(\mathbf{R}))^{3/2}} \left( \frac{|l|}{r} - \frac{2r}{w_0^2} \right). \tag{13}$$

where the Rabi frequency $\Omega(\mathbf{R})$ is calculated at $z = 0$ from Eq.(6). In Fig.2 we present the field lines which correspond to this artificial magnetic field.

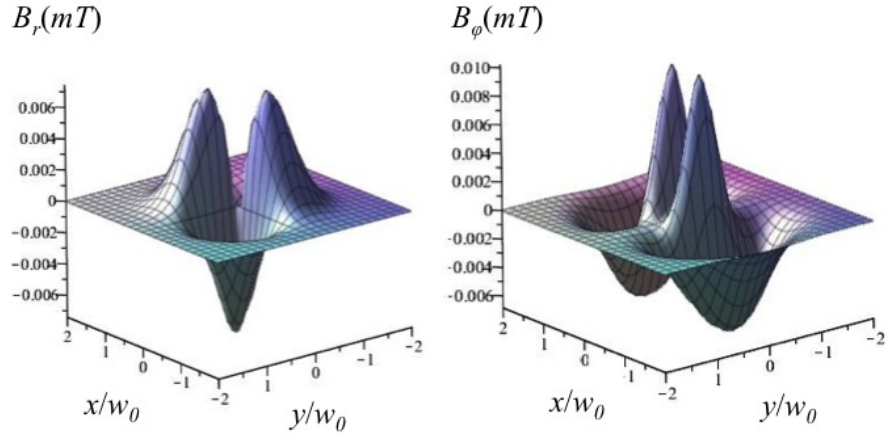

Figure 1. Artificial magnetic field radial and azimuthal components for a two-level atom (transition $6^2S_{1/2} - 6^2P_{3/2}$ in a $^{133}$Cs atom). The atom is irradiated by a Ferris light field with $l = \pm 1$, a beam waist $w_0 = 5\mu m$, while $\Omega_0 = 10\Gamma$ and $\delta = 100\Gamma$

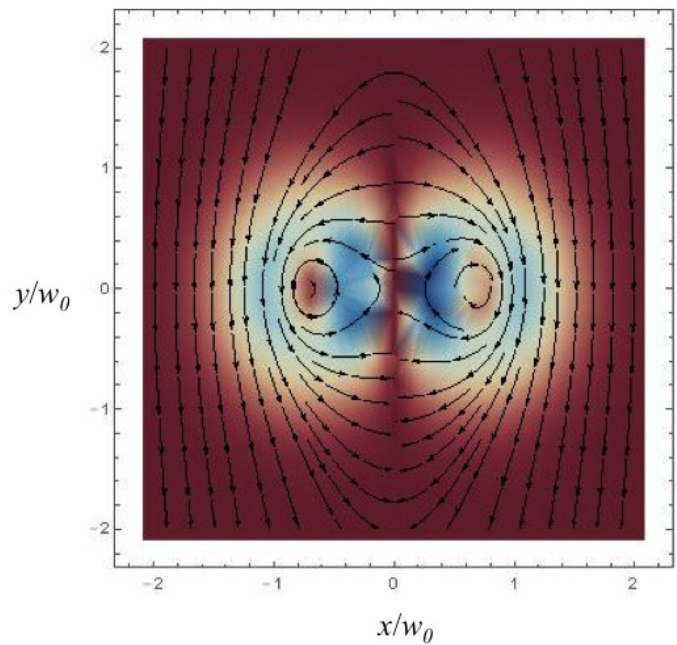

Figure 2. The field lines which correspond to the magnetic field presented in Fig.1

As we clearly see the magnetic field components have a cylindrical symmetry and relatively weak values. The values of the components depend on several parameters like the beam helicity, the beam waist, the detuning and the wave number. The dependence on the beam helicity and beam waist though is not obvious at first sight. First we must recall that the Rabi frequency $\Omega_0$ is related to beam intensity $I = P/\pi w_0^2$, where $P$ is the beam power. As we know for a two-level atom $\Omega_0 = \Gamma\sqrt{I/2I_S}$ where $I_S$ is the so called saturation intensity [23]. Obviously an increase in the beam waist gives a lower Rabi frequency. As the Rabi frequency becomes lower the magnetic field tends to zero. Similarly, as we concern an increased beam helicity, for given power, the beam energy is spread in a larger cross-sectional area, thus the intensity of the beam becomes weaker and so the Rabi frequency and the magnetic field.

The magnetic fields are not the only artificial fields that are created in our case. We have also the creation of a static artificial electric field given by [16]:

$$\mathbf{E}(\mathbf{R}) = -\frac{1}{q}\nabla V. \tag{14}$$

The resulted electric static field has only radial and azimuthal components given by:

$$E_r = -\frac{\hbar^2}{8Mq}\frac{\Omega^2(\mathbf{R})}{\delta^2}\left\{2\left(\frac{|l|}{r}-\frac{2r}{w_0^2}\right)\left[\left(\frac{|l|}{r}-\frac{2r}{w_0^2}\right)^2+\frac{l^2}{r^2}\tan^2(l\phi)+k^2\right]+\right.$$
$$\left.\left[2\left(\frac{|l|}{r}-\frac{2r}{w_0^2}\right)\left(-\frac{|l|}{r^2}-\frac{2}{w_0^2}\right)-\frac{2l^2}{r^3}\tan^2(l\phi)+\frac{2k^2r}{z_R^2}+2k^2\frac{\Omega^2(\mathbf{R})}{\delta^2}\left(\frac{|l|}{r}-\frac{2r}{w_0^2}\right)\right]\right\} \tag{15}$$

$$E_\phi = -\frac{\hbar^2}{8Mqr}\frac{\Omega^2(\mathbf{R})}{\delta^2}\tan(l\phi)\times$$
$$\left\{-2l\left[\left(\frac{|l|}{r}-\frac{2r}{w_0^2}\right)^2+\frac{l^2}{r^2}\tan^2(l\phi)+k^2\right]+\frac{2l^2}{r^2}\sec^2(l\phi)+2lk^2\frac{\Omega^2(\mathbf{R})}{\delta^2}\right\}. \tag{16}$$

In Figs. 3, we present the spatial dependence of the electric field components when a two-level Cs atom (transition $6^2S_{1/2} - 6^2P_{3/2}$) interacts with a Ferris wheel light field with $l = \pm2$, a beam waist $w_0 = 5\mu m$, while $\Omega_0 = 10\Gamma$ and $\delta = 100\Gamma$. In Fig.4 we present the corresponding field lines.

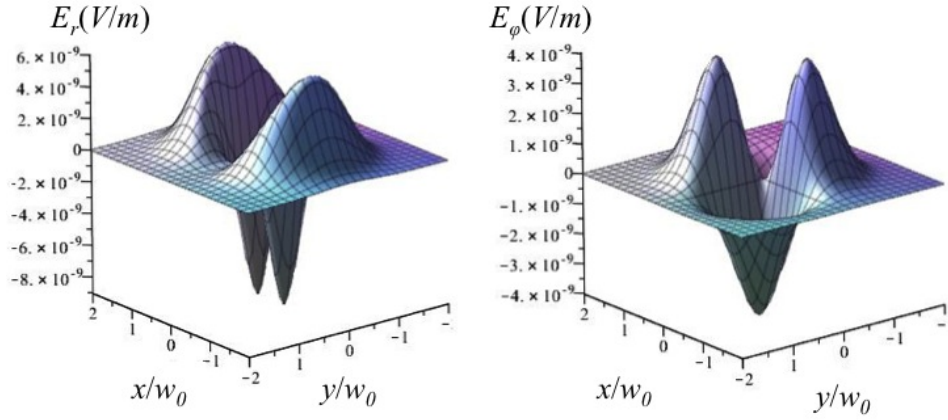

Figure 3. Artificial electric field radial and azimuthal components for a two-level atom (transition $6^2S_{1/2} - 6^2P_{3/2}$ in a $^{133}$Cs atom). The atom is irradiated by a Ferris wheel light field with $l = \pm 2$, a beam waist $w_0 = 5\mu m$, while $\Omega_0 = 10\Gamma$ and $\delta = 100\Gamma$

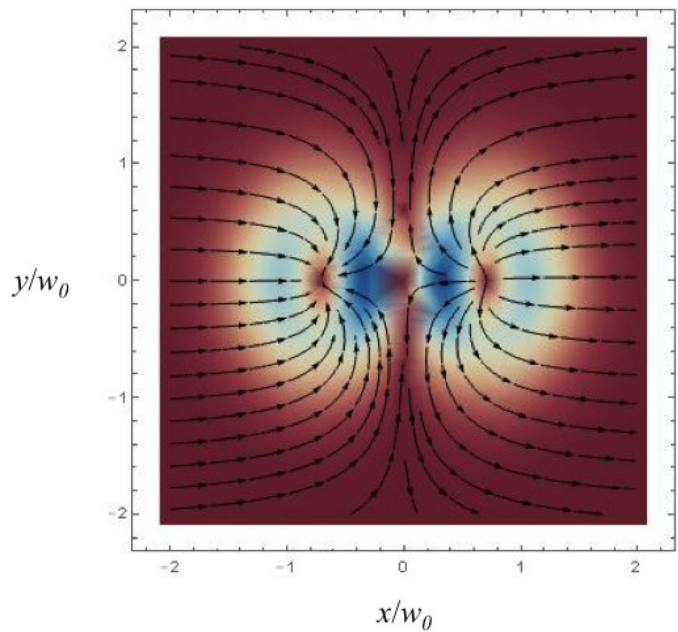

Figure 4. The field lines which correspond to the electric field presented in Fig.3

## 3. Artificial Gauge Fields generated by a rotating Ferris Wheel Light Field

When the frequencies $\omega_1$ and $\omega_2$ of the two superimposed GL beams are slightly different the intensity pattern of the optical Ferris light field is rotating and the Rabi frequency is given by [15]:

$$\tilde{\Omega}(\mathbf{R}, t) = \Omega(\mathbf{R}) \cos(l\phi - \Delta\omega t/2), \tag{17}$$

where $\Delta\omega = \omega_1 - \omega_2$. The formula indicates a rotation angular frequency equal to $\Omega_{rot} = \Delta\omega/2l$. It is also well established that if the interaction is time dependent we also have the generation of an artificial time dependent electric field given by [16]:

$$\mathbf{E}(\mathbf{R}) = -\frac{\partial \mathbf{A}}{\partial t} - \frac{1}{q}\nabla V, \tag{18}$$

where $\mathbf{A}(\mathbf{R})$ is the vector potential associated with the artificial magnetic field in Eq.10, [9], and is given by:

$$\mathbf{A}(\mathbf{R}) = \frac{\hbar}{2}(\cos(\Theta(\mathbf{R}) - 1)\nabla\phi_F(\mathbf{R}). \tag{19}$$

If the time dependent Rabi frequency given in Eq.(17) is inserted in Eqs.(10),(11) and Eq.(18) we are going to get fields that will be functions of the $\cos(l\phi - \Delta\omega t/2)$ indicating a rotation of the fields around the z-axis. Indeed we can show that the beam focus, $z = 0$, the artificial magnetic field has the form:

$$\mathbf{B}(\mathbf{R}, t) = B_r(r, l\phi - \Delta\omega t/2)\hat{r} + B_\phi(r, l\phi - \Delta\omega t/2)\hat{\phi}. \tag{20}$$

This indicates that the interaction simulates a magnetic field which propagates in the $\phi$ direction, so it propagates in a closed loop, with an angular speed $\Omega_{rot}$. Similarly from Eq.(18) we can find the generation of an artificial electric field of the form:

$$\mathbf{E}(\mathbf{R}, t) = E_r(r, l\phi - \Delta\omega t/2)\hat{r} + E_\phi(r, l\phi - \Delta\omega t/2)\hat{\phi} + E_z(r, l\phi - \Delta\omega t/2)\hat{z}. \tag{21}$$

and thus we can state that our scheme is capable of generating artificial gauge electromagnetic fields propagating in closed loops. The direction of the rotation can be reversed by just changing the sign of the relevant beam helicities from $\pm l$ to $\mp l$.

The expressions for the artificial gauge fields components are very lengthy and complicated. However they can get a simpler form if we consider the case of small values for the beam helicity $l$ and a large detuning, such that $\delta >> \Omega$ then the explicit form of these fields, at $z = 0$ is as follows:

$$B_r = \frac{\hbar l k \Omega^2(\mathbf{R})}{2\pi q \delta^2 r} \sin(2l\phi - \Delta\omega t). \tag{22}$$

$$B_\phi = \frac{\hbar k \Omega^2(\mathbf{R})}{\pi q \delta^2} \left( \frac{|l|}{r} - \frac{2r}{w_0^2} \right) \cos^2(l\phi - \Delta\omega t/2) \tag{23}$$

The time-dependent character of the Rabi frequency ensures also the existence of a time-dependent artificial electric gauge field. The expressions for the electric field components (at $z = 0$) are the following:

$$E_r = -\frac{\hbar^2}{8Mq}\frac{\tilde{\Omega}^2(\mathbf{R},t)}{\delta^2}\left\{2\left(\frac{|l|}{r}-\frac{2r}{w_0^2}\right)\left[\left(\frac{|l|}{r}-\frac{2r}{w_0^2}\right)^2+\frac{l^2}{r^2}\tan^2(l\phi-\delta\omega t/2)+k^2\right]+\right.$$

$$\left.\left[2\left(\frac{|l|}{r}-\frac{2r}{w_0^2}\right)\left(-\frac{|l|}{r^2}-\frac{2}{w_0^2}\right)-\frac{2l^2}{r^3}\tan^2(l\phi-\delta\omega t/2)+\frac{2k^2r}{z_R^2}+2\frac{\tilde{\Omega}^2(\mathbf{R},t)k^2}{\delta^2}\left(\frac{|l|}{r}-\frac{2r}{w_0^2}\right)\right]\right\}$$

$$(24)$$

$$E_\phi = -\frac{\hbar^2}{8Mqr}\frac{\tilde{\Omega}^2(\mathbf{R},t)}{\delta^2}\left\{-2l\tan(l\phi-\delta\omega t/2)\left[\left(\frac{|l|}{r}-\frac{2r}{w_0^2}\right)^2+\frac{l^2}{r^2}\tan^2(l\phi-\delta\omega t/2)+k^2\right]+\right.$$

$$\left.\left[\frac{2l^3}{r^2}\tan(l\phi-\delta\omega t/2)\sec^2(l\phi-\delta\omega t/2)-2l\frac{\tilde{\Omega}^2(\mathbf{R},t)k^2}{\delta^2}\tan(l\phi-\delta\omega t/2)\right]\right\}, \quad (25)$$

$$E_z = \frac{\hbar\Delta\omega}{8q}\frac{k\Omega^2(\mathbf{R})\sin(2l\phi-\delta\omega t)}{\delta^2}. \qquad (26)$$

The expressions for the time-dependent fields shows clearly that they rotate in time around the $z$-axis. We demonstrate these rotations in five videos, at the link $https : //www.youtube.com/watch?v = fZBMbaWNX0 cindex = 5 list = UUeGWrUpOYI8KYNihzEkU_dw$, where we present animations of them. The angular frequency of this rotation is, as we mentioned above, $\Omega_{rot} = \Delta\omega/2l$. The question is what are the possible values that this angular frequency may have. The artificial gauge fields which are generated when atoms in the free space interact with coherent light "survive" only for interaction times smaller than $\Gamma^{-1}$ as we have already mentioned. This means that in order to create an artificial electromagnetic field that moves in a closed path we must chose for the optical Ferris field a rotation frequency such that $\Omega_{rot} > \Gamma$. Simultaneously the rotation frequency must not exceed the Rabi frequency $\Omega_0$, so the interaction of the two-level atom with the optical field preserves its adiabatic character. For the values we have in our numerical examples presented in the figures of this paper this means $\Gamma < \Omega_{rot} < 10\Gamma$ which implies a frequency shift between the two beams in the region $2l\Gamma < \Delta\omega < 20l\Gamma$. The rotation frequency is very large for our data thus in the videos we have chosen a small rotation angular frequency to help the viewer to understand the effect.

The magnitude of the artificial fields is very small as it is expected when beams interact with two-level atom [9]. From the plots we see that the radial and artificial magnetic field components are of comparable size. In the case of the electric field we see though that the component $E_z$ is far larger than the other two components. The magnitude of $E_z$ is determined by the frequency shift $\Delta\omega$. As we may see from the videos the ratio of the maximum $E_z$ value to the maximum value of $B_\phi$ is around 20. This, in a very preliminary discussion, indicates a "speed" of propagation around $20m/s$. In other words we have simulated electromagnetic fields that move in an extremely optically dense material. The magnetic fields both in plots and videos are presented in $mT$ units while the electric fields in $V/m$.

## 4. Conclusions

We demonstrated the artificial gauge magnetic and electric fields which are created when a free two-level atom interacts with an optical Ferris wheel light field. We showed that the cylindrical symmetry of the transverse intensity pattern of the Ferris wheel light field is "imprinted" on the artificial gauge fields which exhibit a cylindrical symmetry. We also showed the general formulas for the generated artificial fields in the case of large detuning where the formulas are less complicated and we gave the corresponding plots for the case of specific numerical values of the involved parameters. We demonstrated that for the case where the optical Ferris wheel light field pattern rotates, we get artificial electromagnetic fields that propagate in closed circular paths. We gave videos with animations of these fields.

   One important parameter is the time interval for which these fields do exist. As we said the duration this time interval is determined fully by the excited state transition rate $\Gamma$, which in the case of optical transition in alkalis ensures a very short life time for the artificial gauge fields. To overcome this difficulty, in the case of two levels, it has been proposed to employ lathanides atoms like erbium or dysprosium, which have a more complex electronic structure than alkalis. In these atomic species we can found transitions with very small values of $\Gamma$ [25] thus the artificial fields have a long life time. Also since $\Omega_{rot} > \Gamma$ we can employ rotating Ferris light patterns with very small rotational frequencies and thus the generated artificial electric fields will have far smaller values.

   Finally we point out that magnitude of the artificial fields can be considerably increased when the atoms interact with evanescent fields which are developed near the interface of dielectric media with vacuum as it has been shown [13],[24]. If we use a Ferris wheel light field which will suffer a total internal reflection at a dielectric-vacuum interface we will create an evanescent field which, if it interacts with a two-level atom, the generated artificial fields will be stronger and they will show an elliptical symmetry on their spatial profile. In the case of rotation the fields will propagate on elliptical paths. In this work we demonstrate for first time the possibility of creating artificial e/m fields traveling in closed path. Further investigation is needed on the effects such fields may have on the dynamics of trapped ultracold atoms. Finally we must point out that in nature the paths followed by travelling e/m waves are bend when the fields travel either in media with structured refractive indices or in the vicinity of very massive objects. We need to investigate further how and if our scheme may be used to simulate such physical cases and provide new opportunities to desk-top astronomy.